\documentclass{sigchi}

\CopyrightYear{2020}
\setcopyright{rightsretained}
\doi{}
\isbn{}
\conferenceinfo{arxiv preprint}{}
\acmPrice{}

\usepackage{balance}       
\usepackage{graphics}      
\usepackage[T1]{fontenc}   
\usepackage{txfonts}
\usepackage{mathptmx}
\usepackage[pdflang={en-US},pdftex,pdfpagelabels=false]{hyperref}
\usepackage{color}
\usepackage{booktabs}
\usepackage{textcomp}

\usepackage{microtype}        
\usepackage[all]{hypcap}    
\usepackage{ccicons}          
\usepackage[utf8]{inputenc} 

\def\plaintitle{Maps, Mirrors, and Participants: Design Lenses for Sociomateriality in Engineering Organizations}

\def\emptyauthor{}
\def\plainkeywords{centaur-centered design; design methodology; conceptual design; design models; toolkits; convex optimization; geometric programming}

\makeatletter
\def\url@leostyle{%
  \@ifundefined{selectfont}{
    \def\UrlFont{\sf}
  }{
    \def\UrlFont{\small\bf\ttfamily}
  }}
\makeatother
\def\pprw{8.5in}
\def\pprh{11in}

\definecolor{linkColor}{RGB}{6,125,233}
\hypersetup{%
  pdftitle={\plaintitle},
  pdfauthor={\emptyauthor},
  pdfkeywords={\plainkeywords},
  pdfdisplaydoctitle=true, 
  bookmarksnumbered,
  pdfstartview={FitH},
  colorlinks,
  citecolor=black,
  filecolor=black,
  linkcolor=black,
  urlcolor=linkColor,
  breaklinks=true,
  hypertexnames=false
}

\begin{document}

\title{\plaintitle}

\numberofauthors{3}
\author{%
  \alignauthor{Edward Burnell\\
    \affaddr{Massachusetts Institute of Technology}\\
    \affaddr{Cambridge, MA, USA}\\
    \email{eburn@mit.edu}}\\
  \alignauthor{Priya P. Pillai\\
    \affaddr{Massachusetts Institute of Technology}\\
    \affaddr{Cambridge, MA, USA}\\
    \email{pppillai@mit.edu}}\\
  \alignauthor{Maria C. Yang\\
    \affaddr{Massachusetts Institute of Technology}\\
    \affaddr{Cambridge, MA, USA}\\
    \email{mcyang@mit.edu}}\\
}

\maketitle

\begin{abstract}
When you use a computer it also uses you, and in that relationship forms a new entity of melded agencies, a “centaur” inseparably human and nonhuman. Networks of interaction in an organization similarly form “organizational centaurs”, melding humans, technologies, and organizations into an inseparable sociomateriality. By developing a convex optimization toolkit for conceptual engineering we sought to shape these centaurs. How do organizations go from a high-level concept (“let’s make an airplane”) to a “design”, and in that process what blurred lines between humans and computers bring opportunities for research? We present three metaphors that have been useful lenses across our field sites: considering design models as maps shows how centaurs apportioned legitimacy; looking at design models as mirrors illuminates how they sought validation in their perspectives; and treating design models as participants recognizes their opinions and agency as equivalent to other entities in these centaurs.
\end{abstract}




\section{Introduction}
When you use a computer it also uses you, and in that relationship is formed a new entity of melded agencies, inseparably human and nonhuman. \cite{nardi2012} In these relationships, there can be a feeling of powerlessness, of entrapment in the machinations one is part of. \cite{fleischmann2009, kuijer2018} As designers of human-computer interactions who form such meldings, how can we describe and take responsibility for these consequences of our work?

One framework for doing so is that of ``centaur chess'', a popular name for the ``advanced chess'' Kasparov formed after losing to Deep Blue. \cite{kasparov2010} It centers half-human half-computer {\em centaurs}, focusing on how their parts support each other through the interface between them. \cite{milan2018} Centaur-centered design values both humans and computers for their differences, blurring the lines between their cognitions to seek that which is inextricably of both. \cite{fleischmann2009}

\begin{figure*}
\centering
  \includegraphics[width=\textwidth]{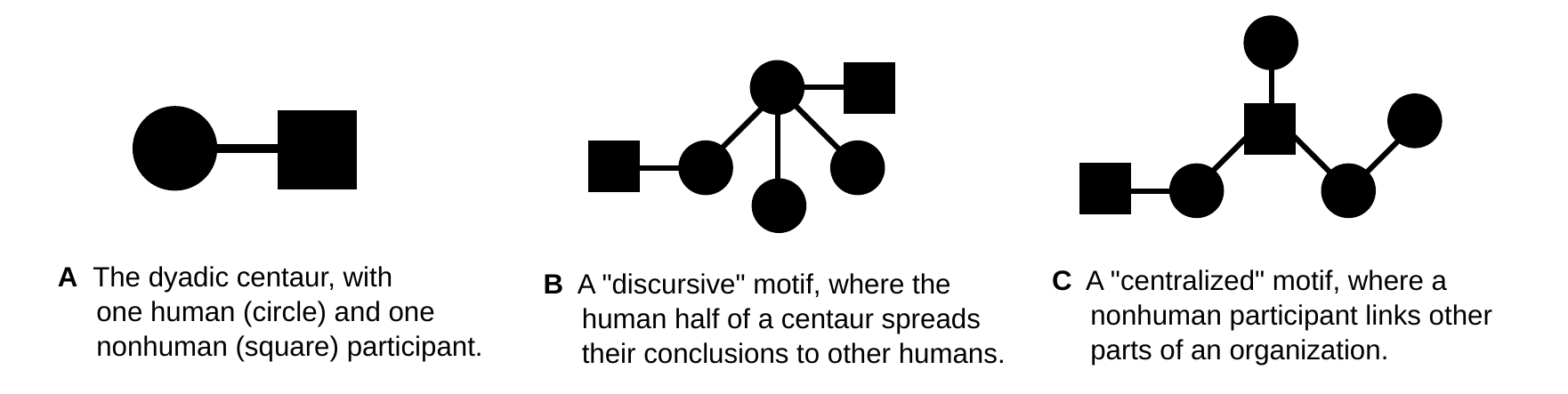}
  \caption{The three organizational centaur motifs used in this work.}~\label{fig:molecules}
\end{figure*}

But the centaur of centaur chess is only a two-node motif in possible arrangements of human-nonhuman networks; more complex motifs, such as those in Figure \ref{fig:molecules}, are also possible and might be framed as {\em organizational} centaurs. The term is used in some sociological literature to mean human-organizational hybrids \cite{alvinius2018}, but we mix it with centaur chess to mean an inextricable blend of human, computer, and organization. Just as a dyadic centaur exists only during moments of human-computer interaction, we can consider organizational centaurs as a constant ``organizing'' practice of human and nonhuman participants. \cite{feldman2011, garud1994} Organizational centaurs represent the many blurred cognitions in the web of practices which constitute an organization. They also obstruct aspiring designers with a thicket of subjectivities, for there is something in the nature of an organizational centaur, emerging as it does from a network of interactions and relationships and so not truly present in a single edge or node, that design and HCI are not always prepared to describe. Organizational centaurs are, however, well described in the literature of sociomateriality, with its focus on the inextricability of humans, organizations, and technologies. \cite{orlikowski2008}

What we sought to design and in this work describe are the organizational centaurs of conceptual engineering design. How do engineering organizations go from a high level concept (``let’s make an airplane'') to a ``design'', and where are the blurred lines between humans and nonhumans in this space that present opportunities for HCI research?

Over the past six years, we have developed a toolkit (``GPkit'') for making engineering design models with an underutilized material: geometric programs, a form of convex optimization. Geometric programs can describe the function, form, and physics of an engineering system with a collection of parameters and algebraic constraints. From the beginning our goal has been to shape organizational centaurs at some of their vital interfaces, and we present here our results and observations from several different field sites. Developing this toolkit required a great respect for both the precise mathematical nature of geometric programs \cite{gpkitchi} and for the knowledges workers at each field site had of their practices \cite{greenwood2006}, for these were two main kinds of cognition we aimed to blur.

Through this process, we arrived at three metaphors useful as lenses through which to see and design for organizational centaurs. Considering design models as {\em maps} points out how centaurs apportioned legitimacy, looking at design models as {\em mirrors} illuminates how centaurs sought to be validated in their perspectives, and treating design models as {\em participants} recognizes their opinions and agency as equivalent to those of other nodes in the centaur. The primary contribution of this work is our observations of the organizational centaurs of our field sites, and how they shaped these lenses.

\section{Background}
During our process we drew upon diverse literatures. As we used each to understand our work, it also entered and changed that work; our understandings of each field site’s practices shaped those practices through communications and the technologies we carried to them. Each ``technology'' we made was a form of reflexivity, an attempt both to understand and to shape what we thought we understood.

\subsection{Design Models}
Human participants in engineering organizations use software "design models" to enumerate parameters of their designs and implement interactions amongst these parameters. Design models are often made from materials like parameterized CAD assemblies (to construct a shape from geometric constraints \cite{leonardi2011, sketchpad}), spreadsheets (to calculate performance \cite{nardi1991, parkin2003}), and ``mathematical programs'' (to take in a desired performance and put out a design that achieves it \cite{mark2002}). These design model materials are all used for dyadic, discursive, and centralized motifs at various frequencies in different organizations.

Design models serve as loci for understanding what will be built, while encoding (and sometimes concealing) decisions on why \cite{robertson2009}. This makes them an important arena for intra-organizational design politics, but just how participants' perspectives clash and coalesce around these models depends also on the motif they are part of \cite{buenza2004, leonardi2011}. Design models express their agency both by shaping the motif and, within a motif, by determining their outsiders and insiders, spectators and maintainers, and formal and informal power structures \cite{greenwood2006, leonardi2011}.

As design models are a clear nonhuman presence in the blurred cognitions of an engineering organization, toolkits for making design models are a powerful tool for shaping and observing organizational centaurs.

\subsection{GPkit as a Toolkit}
Toolkits have been defined by Greenberg as providing a ``vocabulary and set of building blocks'' which ``give people a `language' to think about these new interfaces, which in turn allows them to concentrate on creative designs.'' \cite{Greenberg2007}. Ledo, et al. \cite{ledo2018} extended this to define toolkits as ``generative platforms designed to create new interactive artifacts, provide easy access to complex algorithms, enable fast prototyping of software and hardware interfaces, or enable creative exploration of design spaces''. We have done several of these with GPkit, giving engineers the ability to rapidly build and improve interactive explorations of convex design spaces. GPkit is thus an ``artifact contribution'', where ``new knowledge is embedded in and manifested by artifacts and the supporting materials that describe them'', as demonstrated via the case studies and observations above \cite{ledo2018}.

\subsection{GPkit as a User Interface System}
This work also follows in the footsteps of other user interface systems \cite{olsen2007}. GPkit falls under each of the three limitations of usability experiments, being a tool for expert users doing non-standard tasks over a time period of months or years. As with previous papers using this framework \cite{houben2015} we consider GPkit in Olsen's framework for User Interface Systems \cite{olsen2007}: it introduces a new task to a preexisting group of users (engineering designers) in a preexisting situation (early-stage design), but (unlike in Olsen), GPkit also seeks to change that situation, to alter the ways in which computers are used in early-stage design.

\begin{table*}
  \centering
  \begin{tabular}{l r l l}
    \textbf{Field Site} & \textbf{Modelers} & \textbf{GPkit Design Models} & \textbf{Typical Motif}\\
    Class project with (under)graduate students
        & 16 &  Electric ``air taxi'' service
        & {\em Dyadic}\\
   Aerospace R\&D firm of 100-500 employees
        & 12 & Costing and sizing for airplane projects
        & {\em Discursive}\\
    University researchers
        & 10 & Passenger jets, solar planes, other vehicles
        & {\em Discursive} \\
    Industry-government-university initiative
        & 10 & Hybrid propulsion topologies
         & {\em Dyadic}\\
    Transportation startup of 150-250 employees
        & 7 & Pre-production system engineering
        & {\em Centralized}\\
    Class project with (under)graduate students
        & 3 &  Gas-powered high-altitude surveillance drone
        & {\em Centralized}\\
  \end{tabular}
  \caption{Overview of field sites. The ``Modelers'' column is an approximation of active weekly modelers during the period of study.}
  \label{tab:sites}
\end{table*}

\subsection{GPkit as User Systems Architecture}
GPkit sets up ``paths of least resistance'' \cite{myers2000, olsen2007} for explaining design codes; this results in models that can be more amenable to iterative improvement than alternatives, allowing users to derive related solutions with reduced development viscosity and build on common infrastructure. Convex programs have been inaccessible to most engineering organizations, but by making them accessible, GPkit helps increase the scale of their design models from dozens of free variables to thousands. GPkit thus lowers the threshold and raises the ceiling for design codes. The ``Moving Targets'' present in any use-motivated work \cite{myers2000} have been a fuel for the development of GPkit; we have encouraged participants at our field sites to change their goals and expectations during use so they might adopt previously untenable motifs and practices.

\subsection{Sociomateriality and its influences}
That organizations are shaped by their artifacts has been acknowledged academically since Marx, if not earlier, but Latour’s posthumanist analysis of this, focusing on the ever-shifting alliances between human and material agencies, provides a framework for considering the human and nonhuman as inseparable. \cite{latour1992} Structuration theory opposes this, claiming technology changes organizations because individuals use it as an occasion for reshaping their organization, which then reshapes them and so on until it is humans and organizations whose agencies are inseparable. \cite{barley1986}

Sociomateriality attempts to synthesize both perspectives, holding individual, organizational, and material agencies as fundamentally inseparable and occurring only in concert. \cite{beane2015, introna2011, orlikowski2008, orlikowski2014} This presents challenges for many methods of description and analysis.

\section{Methods}
One of the main difficulties sociomateriality can  present to a designer is its epistemology, its paradigm of where truths come from. Sociomateriality is interpretivist, and considers reality to be socially constructed. \cite{burrell1979} This paradigm is fundamentally different from the positivism common in design and HCI, which consider reality as existing independently, such that subjective perspectives are only limitations of an objective truth. \cite{duarte2016} Interpretivism centers context and criticality, de-centers objectivity and generalizability, and strives for a rigorous humility. If the purpose of much HCI and design research has been to say ``this is the technology we made'', how do we accept not being the sole, or even the primary, authorities on that? As developers we tend to see ourselves at the center of ``our'' technology’s use, and would like to take responsibility for all of our field sites’ positive and interesting outcomes while ignoring the negative, unethical, or uninteresting ones. To address the limitations of our perspective, we describe in the rest of this paper  what we’ve made in the form of vignettes which are explicit about the roles we played, and try to include positive, negative and ambiguous outcomes.

Additionally, interpretivist researchers often prefer to describe what they see rather than prescribe what should be. Design and HCI, on the other hand, can seem at times more comfortable prescribing solutions than describing observations of prototypical use. But this confidence that technologies can be improved, that designers can learn to grapple with ``wicked problems'' seems to us both valuable and necessary. \cite{kling1991, rittel1973} As part of our care to incorporate our subjectivities into this work, it seems to us important to say what we think should be done.

In an attempt to ``handle the balance between [being] a member of the implementation team and an observer'' \cite{schultze2000} in our field sites (see Table \ref{tab:sites}), this text tends towards the confessional. So: we derived, we coded, we documented, we maintained; we demonstrated, we helped, we supported, we explained; in the end, we are left with some kind of technology expressed variously across field sites, and how we chose to understand its actors and actions.

\section{Results}
To describe our toolkit’s shaping of organizational centaurs we use three guiding metaphors \cite{morgan1980}. Considering design models as maps points out how these centaurs apportioned legitimacy, looking at design models as mirrors illuminates how centaurs sought to be validated in their perspectives, and treating design models as participants recognizes their opinions and agency in the design process.

\subsection{Design Models as Legitimizing Maps}
A common metaphor in design is ``exploration'', sometimes in frankly colonialist promises of the ``untapped territories'' accessible with new technology. Design models serve these explorers as maps permitting navigation in lands of possibility; what or who are represented by them shapes the process of design. Kaplan’s study of a telecommunications design process call such determinings of legitimacy, often through the exclusion of alternative perspectives, ``cartography''. \cite{kaplan2011}

\begin{figure*}
  \centering
  \includegraphics[width=\textwidth]{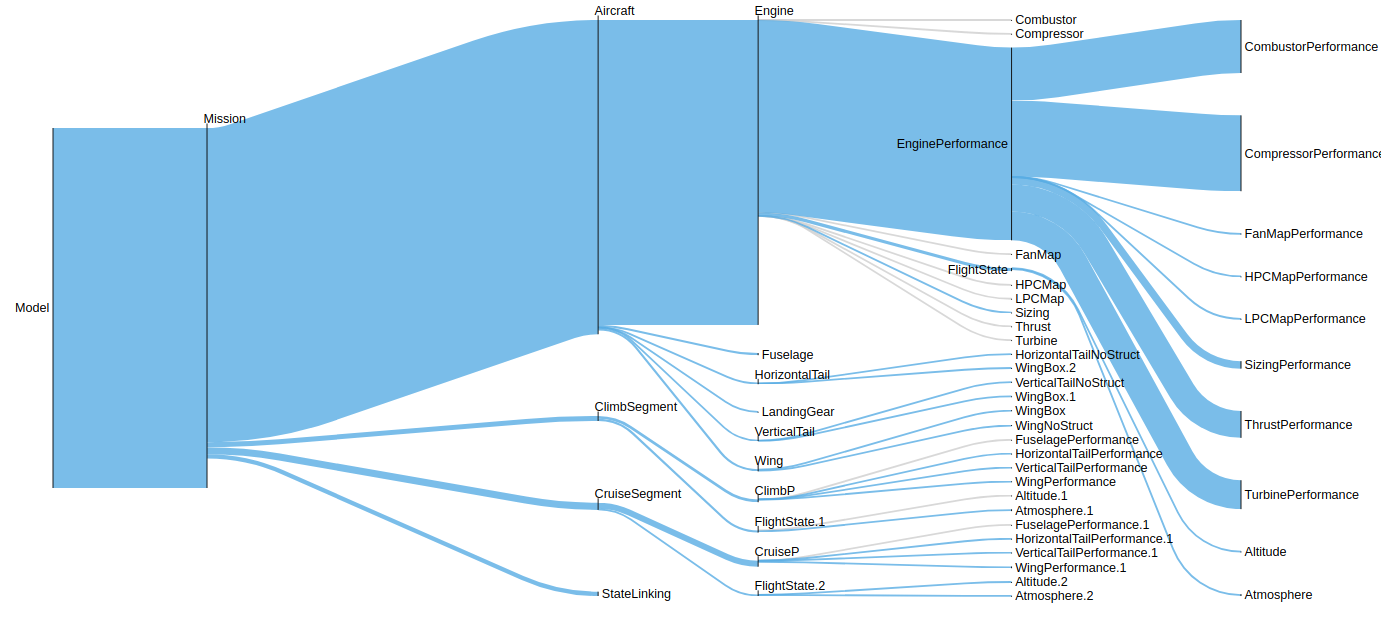}
  \caption{Sankey diagram displaying a model's constraint tree (left is towards the root) and the sensitivity of each named constraint-set (thicker is more sensitive). Discontinuous increases in thickness at a particular object indicate the sensitivity of the unnamed constraints inside it.}
  ~\label{fig:sankey}
\end{figure*}

As a toolkit designer, cartography forms an essential and powerful part of how you shape organizations by your work, and you are implicated in this use. For you make materials of which such maps are made and improved, from early sketches of ``here be dragons'' to later ones so fine and detailed they may seem a substitute for reality itself.

Cartography is fraught: maps always tempt mistaking them for the territory they represent. After all, the legitimacy of a map often rests in its claim to show reality. So too with engineering design models, many of which are called ``simulations'' and used to support design decisions by their claims to approximate the ``real world''. Because of this rhetoric of realism, simulations are caught in the paradox, of {\em wanting} to perfectly capture the detail of the world but {\em needing}, to be practical, to be a compressed and portable version of it.  Instead of space (as with the map of Borges’ story, that grew so large as to be laid out one to one over the land it represented \cite{borges1998}), the resources simulations expand into are time and computation; it can take as long (and as much electricity) to do a detailed computational fluid dynamics simulation as to make a physical prototype and test it in a wind tunnel. In the face of this tradeoff modelers make maps of their maps, known as ``surrogate models'', to quickly estimate the approximations of a slow simulation, helping it fit better into the timescales of a design process. Modelers also make ``low fidelity'' models directly. Despite deviating from it, both surrogation and fidelity still enshrine their ideal of truth as the one-to-one map.

Simulations are the baseline design model, and with them the assumption that the model’s legitimacy derives from its proximity to an objective reality. With GPkit, we sought  to encourage design models to focus instead on the subjective reality of a particular organization and project.

In one memorable incident, this deviation from the simulationism led to a GPkit model being seen as an epistemological affront. Some design participants at one of our field sites asked their primary modeler to run an analysis on the effects of changing one part for a worse version. As later relayed to us by that modeler, the resulting speeds proved better than they expected; they asked the modeler why and were taken aback to learn the design model had reconfigured the entire transportation system to accommodate this worse part. With some emotion, they replied ``that’s unfair!'' After a pause for thought, they added ``because it’s unscientific!'' They felt marginalized by the nature of this design model; parameters they’d thought of as necessarily inputs were represented as outputs. The emotion of their response came in some part from the unexpectedness of their marginalization; design models they’d interacted with before had constructed a virtual reality for designs where each part was ``manufactured'' independently, then set in motion, and so this model that did not do so seemed illegitimate. By itself, this might not have been an issue; but in the centralized centaur motif common in this organization, they perceived a threat of the model delegitimizing them and so seized upon their connections in the present discursive motif and tried to exclude the model for being unscientific. In this case, the design model had more power than the affronted, and so, other than reassurance that they were not being delegitimzed, their protests were unaddressed and ultimately forgotten. But how did this design model acquire such legitimacy?

One straightforward way acquire legitimacy is to confirm or further other’s opinions, as with the Dieppe world maps made for French royalty showing vast fictitious islands available for colonizing. \cite{brownstein2013} A researcher at a different field site once came to us with frustrated certainty that the jet engine component of their design model was ``too brittle'': ``The plane is being entirely designed around the engine! [...] changing the plane model doesn't change the optimal engine at all; changing the optimal engine model completely changes the plane. It's like an engine with a stick attached.'' Part of their frustration was that, while {\em they} had become certain of this by solving for various points, they lacked confidence in their ability to convince their colleague who developed the engine model. We introduced into GPkit a visualization of the relative importance of each submodel to the optimal solution, and produced for them the exact map of Figure \ref{fig:sankey}. It was a vindication: here, clearly labeled, they saw confirmation of their design model's sensitivity to the engine to the exclusion of the rest of the airplane. It turned what had seemed like an insubstantial sensation into something they were confident of presenting to others. 

Another time, a firm hired a researcher explicitly to use GPkit for validating (but not critiquing or modifying) previous decisions on a rather novel airplane design; a direct request for a map whose only purpose was legitimation through confirmation. Such a request was conceivable because, as modelers well know, any search for reality in a design model can be prone to confirmation bias, wishful thinking, and typographical errors. What is produced is always partly a fantasy. Early in a gas-powered surveillance drone class project at one of our field sites, a design model found a curious way of flying; accelerating as it increased in altitude, but unexpectedly keeping the same speed even as it spent fuel and became lighter. The culprit turned out to be a single variable, the Reynolds number (a standard engineering characterization of fluid flow), which had been incorrectly specified as a scalar rather than as a vector. The model was thus, in following this specification, finding an optimal single Reynolds number it could maintain across all modes of flight, making the airplane's speed a function only of its altitude. Years later, the image of this plane striving to achieve a fantastically consistent Reynolds number is still considered extremely comedic by some of the teaching faculty. Of course, as the toolkit designer, the way fantasies are constructed through your work is part of your design, and something that you may feel implicated in. Our first author, in providing support to this class, developed a syntax to make such fantastic oddities easier to avoid. But they had not anticipated assisting a project whose stated surveillance target of disaster-response mapping was probably at best a secondary objective for the ``client'' funding the class, and felt caught between the fantasies of students, staff, and client and their own professional, pedagogical, and ethical responsibilities.

As maps, design models can be taken over-literally, and act to change participant’s perceptions of truth despite being made from fantasies and temporary perspectives. But design models also provide a site and opportunity for participants to reflect on and develop their perspectives; they are maps, but they are also mirrors.

\subsection{Design Models as Validating Mirrors}
Design models reveal modelers to themselves, with various distortions. This was apparent in the affront above; those affronted looked into the plots and saw the vehicle they were designing travel, stop to have one part replaced, then travel again. Their shock at realizing the train had reconfigured itself around the second part was akin to the dysmorphia of seeing your reflection reconfigure itself in the mirror. \cite{schon1983} As a toolkit designer, seeing design models as mirrors thus requires a consideration of the surfaces they present for reflection.

For instance, the mathematics of geometric programs optimization allowed for the kind of flexibility between inputs/outputs, requirements/objectives, and statics/dynamics that human designers sometimes have, and this was essential to the epistemological affront above. Participants’ attempt to define simulation as a fairer, more scientific comparison came from seeing the swapping of the dozen or so parameters of the part as an appropriately minimal change, since thousands of other parameters concerning the construction of the system were held still, ``controlled''. But for optimization algorithms, an engineering system is a network of constraints which interlink to determine the values of those thousands of parameters. When the small number of constraints relating to that part were changed, the model let thousands of construction parameters shift because to do otherwise would require changing the hundreds of other constraints held ``controlled''. To those affronted, this was a funhouse distortion, but to the modeler it was a reflection of their perspective on the nature of the vehicle’s design.

Particularities of convex optimization programs are perhaps where our advice is both most technical and least generalizable, though we encourage the reader to recognize qualities like these in technologies you shape. Another property of convex optimization programs is that they have relatively few and obvious assumptions; thousands of parameters are often determined by hundreds of individually-straightforward algebraic constraints and a choice of objective function. This efficiency is implicated in both the affront and the constant-Reynolds comedy above. Despite their size, convex programs respond on the order of seconds, speeding human processes of intuition-building and reflection. The convex geometric programs with which we worked also provide a ``dual solution'', giving in addition to optimal parameter values the sensitivity that result to each constraint. Knowledge of such sensitivities are something human modelers tend to build, although they sometimes (as in the ``engine with a stick attached'') can seem insubstantial without a method of reflection. As modelers come to know this surface of their models we see cognition blur. The confirming sensitivity map was a frustrated thought from the human side and a contextless knowledge from the nonhuman one; as a thought it exists between them.

For the organizational centaurs of conceptual engineering, design also involves a great deal of {\em group} reflection. Discursive centaurs often occur in the practice of ``design reviews'', in which design participants present their perspectives and decisions to other participants. At one aerospace R\&D field site, an engineer said they used GPkit primarily for ``interactive presentations of novel design spaces''. Because convex programs have fewer assumptions, it is easier to validate them; because they’re fast, it’s possible to resolve them live, and because solutions are explainable through their dual, participants in the audience (who often have more opinions on the expected tradeoffs of a design than on its exact parameter values) can more quickly critique a new model.

One property of convex programs which reflected discursive and centralized centaurs much more than dyadic ones: their requirement for ontological specificity through explicitly coded assumptions. Baldwin famously noticed that modular decomposition of a design often corresponds to the organizational structure of its designers, \cite{baldwin2000} but what occurs when that structure is made more legible to participants? We have seen in several field sites conversations around convex design models begin incorporating explicit discussions of such ontologies. In particular we have heard modellers increasingly refer to where constraints ``belong''. This was done explicitly in ``\#TODO'' comments apologizing for a constraint's misplacement, in ``\#NOTE'' comments justifying its presence, and in design reviews (``does the fuel burn constraint belong here [with the fuel model] or there [with the airplane drag constraints]?''). It was also done implicitly, ``belonging'' represented by the clustering of constraints into named categories. Constraints were generally described as ``belonging'' either to categories arising from a hierarchy of the design's physical layout (\texttt{wing}, \texttt{fuselage}) or from disciplinary hierarchies (\texttt{aerodynamics\_eqs}, \texttt{structural\_eqs}) common in aerospace engineering. When we let modelers represent such clusters as custom code objects (for example, an Airplane instance containing Wing and Fuselage instances), they began clustering via object-oriented ontologies. Although most modelers had not written prior object-oriented code, adoption of this feature was surprisingly rapid; the object-oriented framework seemed to fit their experience and intuition. Intriguingly, different field sites designing similar airplanes often came up with incompatible ontologies. This organizing of constraints into clusters seems not to be a convergence onto some ``ideal'' ontology, but rather a divergence of dialects, saying similar things with different pronunciations. The design models of each organization were active in the development of this language; we often heard modellers say the model they worked with ``wanted'' constraints to be in a particular place, especially if its opinions differed from theirs.

\subsection{Design Models as Opinionated Participants}
This leads us to consider design models as participants in the design process. Set-based design holds that design processes are a continuous narrowing (through the expertise of workers) of an initially vast design space, ending when only a single design remains. \cite{ghosh2014} On an organizational level, set-based design sees the role of participants as being to set constraints in the space of possible designs (such as ``we certainly (won’t be able to/don’t want to) go faster than this''), and so encourages organizations to make it easy for each worker to add constraints, and difficult for others to remove them. In other words, set-based design frames the organization as a mathematical program where each worker is seen as a distinct and self-improving constraint. Each worker is treated as if they were a design model.

Imagine asking ``how do you know your design’s parameters?'' of a human design participant. They may consider themselves the authority on some parameter: whatever they say such a parameter is, it is. They’ll likely recognize coworkers as authorities on other parameters, but they’re also likely to recognize particular design models as authorities on some parameters: ``once we give it beam lengths, the structural analysis software tells us the stiffness.'' The epistemological question of ``how do you know your parameters'' has become a question of locating authority in participants and models. Of course, not all design participants will have consistent answers, and there are ample opportunities for bias in setting constraints and margins. \cite{austin2016} There’s likely to be some contention in who has legitimate authority over particular parameters. There may also be disagreement on whether some parameters exist at all, with some design participants saying they do (``material choice is important!''), some disagreeing (``it’s too early to choose a material!''), and others abstaining or oblivious.

In a set-based design framework, the design model is seen as a dynamic consensus, a coming-together of the actors around it. It can hold the opinions of human participants just as well as those of other design models. This makes it subjective: its role here, instead of mirroring an individual, is to mirror its network, to provide the subjective consensus design space of its particular location. \cite{ghosh2014}

This social consensus can also serve as an enforcer, and this has been important to some applications of set-based design. It was relayed to the authors in conversation that its major benefit to one military organization had been reducing a pattern in which admirals assigned to a project would drastically change its priorities (``no, the most important thing in modern small craft is speed''), scrapping already-achieved design convergence, and then leave for the next admiral to come in and yank the project in yet another direction. Set-based design models prevent this by acting in a way that doesn’t follow the Navy’s regular chain of command; constraints they hold on parameters such as speed may have been established by lower-ranking officers without requiring them to disobey an admiral. That this is the agency of the model, not of the participants who created its constraints, is vital to its power over admirals; the model is less susceptible to the chain of command. Such patterns of authority occur outside the military as well, of course.

As a toolkit designer, working towards the creation of a new design participant requires a kind of boundary-setting. As soon as a design model becomes part of an organization, it is a stranger. It knows things, and has relationships with other design participants, that you do not. You have to visit it regularly to catch up and hear its thoughts, and your knowledge of its construction only takes you so far in these interviews. The design model will never again be something you can fully know. This is, in fact, something to strive for; the level to which you as an outsider can grasp the model’s meaning from your knowledge of its toolkit is also the level to which it does not represent a nuanced local consensus.

\begin{figure}
\centering
  \includegraphics[width=\columnwidth]{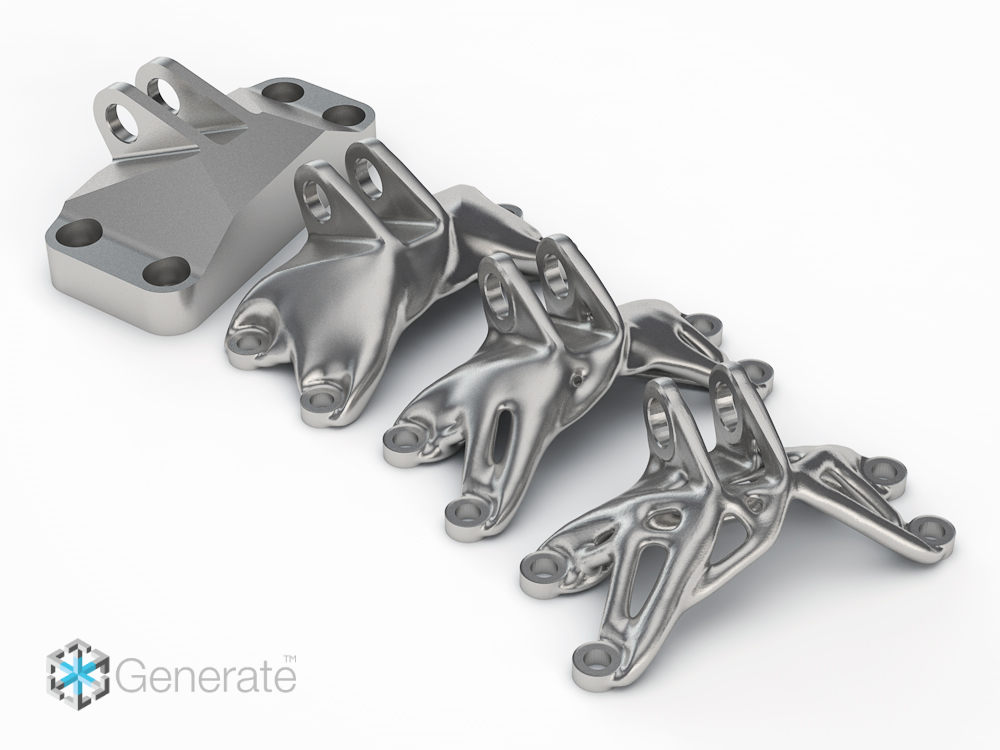}
  \caption{Visualization of a topology optimization algorithm. \protect\cite{topoptfig}} 
  \label{fig:topopt}
\end{figure}

One corollary of seeing design models as participants is to see them as automating replacements for humans (or vice-versa \cite{shestakofsky2017}). This is often seen with topology optimization, a family of structural design algorithms which produce fantastically organic forms like that in Figure \ref{fig:topopt}. \cite{bendsoe1988} The weight-savings achieved by combining this software with skilled human designers provokes a fantasy of replacing those human designers; the characteristic shape of this kind of optimization’s geometries has been called ``the aesthetic of the computer'', when it is really a blurred cognition, the combined aesthetic of the algorithm and its culture of use (the choice of voxel density, shapes, and loading conditions). That is to say, topology optimization’s wonderful results are produced by  centaurs, but these centaurs are often seen only for their nonhuman actors.

\subsection{Case Study: Application to Topology Optimization}
While observations of use can present honest muddlement in the hopes of insight, design lenses work best when they are clear. In this section we work an example of treating the motifs of organizational centaurs as users, and our metaphors as lenses suggesting effective designs.

\subsubsection{Topology Optimization as a Map}
Imagine a tool that attempted to present the fundamentally foamy (variable-density voxel) reality of topology optimization to an organization. How might a topology optimization serve to affront or confirm participants’ perspectives? Topology optimizations already serve well as fantasies, but their fixation on building authority through realism has prevented them from the collaborative possibilities of their visual surrealism, a phenomenon also noted in the creation of CAD models; they seem well suited to a series of provocative ``what-ifs'', and an algorithm that intentionally processed multiple scenarios in parallel might be able to achieve some speed improvements. For a discursive centaur, the ability to show multiple options would allow for an apportioning of relative legitimacy that might increase the legitimacy of the model eventually chosen.

\subsubsection{Topology Optimization as a Mirror}
For a centralized centaur, where participants collaborate on a shared model, the multiple-scenario algorithm mentioned above might mirror their differences in perspective or priorities. For discursive or dyadic motifs we note that topology optimization proceeds through iterated solves converging on a solution; perhaps single iterations could be used as a notion of ``pressures'' on an existing CAD model, or drawn quickly enough that modelers could work more interactively, blurring the cognition of structural design decisions.

\subsubsection{Topology Optimization as a Participant}
How might topology optimization better collaborate with others as an equal participant? How can it keep its fluid aesthetic opinions while becoming a better listener, more attuned to the particular context it is used in? Organizational development of internal ``building blocks'' or templates might aid this, as might (for a centralized motif) a system of incrementally updating the topology-optimized parts as the rest of the geometry changes. The question of how topology optimization might grow from working with experienced structural engineers (as it has from working with experienced structural-analysis researchers) is an important one here, but would require creating prototypes and releasing them in a field site to become strangers.

\section{Conclusion}
In this work we have described our observations of organizational centaurs in engineering organizations and our attempts to shape them through developing a toolkit for design models. Seeing design models shape legitimacies as maps, reflect particularities of thought as mirrors, and join design processes as opinionated participants shaped our observations, and we’ve attempted to playfully and confessionally present our perspective by weaving metaphors and vignettes into a thick subjectivity. But just describing is only half our goal; we hope also to show the benefits of actively using these metaphors as lenses during a design process. Designing for these more-than-human organizational centaurs can be fraught, but we think these lenses provide a humble clarity; this is necessary as we as designers begin to take responsibility for the implications and melded agencies of our designs.

\balance{}

\bibliographystyle{SIGCHI-Reference-Format}
\bibliography{zotero}


\begin{thebibliography}{00}


\ifx \showCODEN    \undefined \def \showCODEN     #1{\unskip}     \fi
\ifx \showDOI      \undefined \def \showDOI       #1{{\tt DOI:}\penalty0{#1}\ }
  \fi
\ifx \showISBNx    \undefined \def \showISBNx     #1{\unskip}     \fi
\ifx \showISBNxiii \undefined \def \showISBNxiii  #1{\unskip}     \fi
\ifx \showISSN     \undefined \def \showISSN      #1{\unskip}     \fi
\ifx \showLCCN     \undefined \def \showLCCN      #1{\unskip}     \fi
\ifx \shownote     \undefined \def \shownote      #1{#1}          \fi
\ifx \showarticletitle \undefined \def \showarticletitle #1{#1}   \fi
\ifx \showURL      \undefined \def \showURL       #1{#1}          \fi

\bibitem{alvinius2018}
{Aida Alvinius} {and} {Gerry Larsson}. 2018.
\newblock \showarticletitle{Applying the {Organizational} {Centaur} {Theory} on
  {Boundary} {Spanners} in {Demanding} {Situations}}.
\newblock {\em Crisis Management - Theory and Practice\/}  {1} (March 2018),
  20.
\newblock
\showDOI{%
\url{http://dx.doi.org/10.5772/intechopen.74712}}


\bibitem{austin2016}
{Jesse Austin-Breneman}, {Bo~Yang Yu}, {and} {Maria~C. Yang}. 2016.
\newblock \showarticletitle{Biased information passing between subsystems over
  time in complex system design}.
\newblock {\em Journal of Mechanical Design\/} {138}, 1 (2016), 9.
\newblock


\bibitem{baldwin2000}
{Carliss~Young Baldwin}, {Kim~B. Clark}, {and} {Kim~B. Clark}. 2000.
\newblock {\em Design rules: {The} power of modularity}. Vol.~1.
\newblock MIT Press, Cambridge.
\newblock


\bibitem{barley1986}
{Stephen~R Barley}. 1986.
\newblock \showarticletitle{Technology as an occasion for structuring:
  {Evidence} from observations of {CT} scanners and the social order of
  radiology departments}.
\newblock {\em Administrative Science Quarterly\/}  {31} (1986), 78--108.
\newblock


\bibitem{beane2015}
{Matt Beane} {and} {Wanda~J Orlikowski}. 2015.
\newblock \showarticletitle{What difference does a robot make? {The} material
  enactment of distributed coordination}.
\newblock {\em Organization Science\/} {26}, 6 (2015), 1553--1573.
\newblock


\bibitem{bendsoe1988}
{Martin~Philip Bendsoe} {and} {Noboru Kikuchi}. 1988.
\newblock \showarticletitle{Generating optimal topologies in structural design
  using a homogenization method}.
\newblock {\em Computer Methods in Applied Mechanics and Engineering\/}  {71}
  (1988), 28.
\newblock


\bibitem{borges1998}
{Jorge~Luis Borges}. 1998.
\newblock \showarticletitle{On {Exactitude} in {Science}}.
\newblock In {\em Collected {Fictions}}. Penguin, New York.
\newblock


\bibitem{brownstein2013}
{Daniel Brownstein}. 2013.
\newblock Java {La} {Grande}.
\newblock   (May 2013).
\newblock
\showURL{%
\url{https://dabrownstein.com/2013/05/12/java-la-grande/}}


\bibitem{buenza2004}
{Daniel Buenza} {and} {David Stark}. 2004.
\newblock \showarticletitle{Tools of the trade: the socio-technology of
  arbitrage in a {Wall} {Street} trading room}.
\newblock {\em Industrial and Corporate Change\/} {13}, 2 (2004), 369--400.
\newblock


\bibitem{gpkitchi}
{Edward Burnell}, {Nicole~B. Damen}, {and} {Warren Hoburg}. 2020.
\newblock \showarticletitle{{GPkit}: {A} {Human}-{Centered} {Approach} to
  {Convex} {Optimization} in {Engineering} {Design}}. In {\em Conference on
  {Human} {Factors} in {Computing} {Systems} ({CHI})}. Association for
  Computing Machinery, Honolulu, 12.
\newblock
\showDOI{%
\url{http://dx.doi.org/10.1145/3313831.3376412}}


\bibitem{burrell1979}
{Gibson Burrell} {and} {Gareth Morgan}. 1979.
\newblock \showarticletitle{Assumptions {About} the {Nature} of {Social}
  {Science}}.
\newblock In {\em Sociological paradigms and organisational analysis:
  {Elements} of the sociology of corporate life}. Heinemann, London, 38.
\newblock


\bibitem{duarte2016}
{Emanuel~Felipe Duarte} {and} {M.~Cecília~C. Baranauskas}. 2016.
\newblock \showarticletitle{Revisiting the {Three} {HCI} {Waves}: {A}
  {Preliminary} {Discussion} on {Philosophy} of {Science} and {Research}
  {Paradigms}}. In {\em Brazilian {Symposium} on {Human} {Factors} in
  {Computing} {Systems}}. Association for Computing Machinery, São Paulo,
  1--4.
\newblock


\bibitem{feldman2011}
{Martha~S. Feldman} {and} {Wanda~J. Orlikowski}. 2011.
\newblock \showarticletitle{Theorizing practice and practicing theory}.
\newblock {\em Organization science\/} {22}, 5 (2011), 1240--1253.
\newblock


\bibitem{fleischmann2009}
{Kenneth~R. Fleischmann}. 2009.
\newblock \showarticletitle{Sociotechnical interaction and cyborg–cyborg
  interaction: transforming the scale and convergence of {HCI}}.
\newblock {\em The Information Society\/} {25}, 4 (2009), 227--235.
\newblock


\bibitem{garud1994}
{Raghu Garud} {and} {Michael~A Rappa}. 1994.
\newblock \showarticletitle{A socio-cognitive model of technology evolution:
  {The} case of cochlear implants}.
\newblock {\em Organization science\/} {5}, 3 (1994), 344--362.
\newblock


\bibitem{ghosh2014}
{Sourobh Ghosh} {and} {Warren Seering}. 2014.
\newblock \showarticletitle{Set-based thinking in the engineering design
  community and beyond}. In {\em International {Design} {Engineering}
  {Technical} {Conferences} and {Computers} and {Information} in {Engineering}
  {Conference} ({IDETC}/{CIE})}. American Society of Mechanical Engineers,
  Buffalo, 13.
\newblock


\bibitem{Greenberg2007}
{S. Greenberg}. 2007.
\newblock \showarticletitle{Toolkits and interface creativity}.
\newblock {\em Multimedia Tools and Applications\/} {32}, 2 (2007), 139 --159.
\newblock
\showDOI{%
\url{http://dx.doi.org/10.1007/s11042-006-0062-y}}


\bibitem{greenwood2006}
{Davydd~J Greenwood} {and} {Morten Levin}. 2006.
\newblock {\em Introduction to {Action} {Research}: {Social} {Research} for
  {Social} {Change}}.
\newblock SAGE Publications, Thousand Oaks.
\newblock


\bibitem{houben2015}
{Steven Houben} {and} {Nicolai Marquadt}. 2015.
\newblock \showarticletitle{Watchconnect: {A} toolkit for prototyping
  smartwatch-centric cross-device applications}. In {\em Conference on {Human}
  {Factors} in {Computing} {Systems} ({CHI})}. Association for Computing
  Machinery, Seoul, 1247--1256.
\newblock


\bibitem{introna2011}
{Lucas~D. Introna} {and} {Niall Hayes}. 2011.
\newblock \showarticletitle{On sociomaterial imbrications: {What} plagiarism
  detection systems reveal and why it matters}.
\newblock {\em Information and Organization\/} {21}, 2 (2011), 107--122.
\newblock


\bibitem{kaplan2011}
{Sarah Kaplan}. 2011.
\newblock \showarticletitle{Strategy and {PowerPoint}: {An} inquiry into the
  epistemic culture and machinery of strategy making}.
\newblock {\em Organization Science\/} {22}, 2 (2011), 320--346.
\newblock


\bibitem{nardi2012}
{Victor Kaptelinin} {and} {Bonnie Nardi}. 2012.
\newblock {\em Activity theory in {HCI}: {Fundamentals} and reflections}.
\newblock Morgan \& Claypool Publishers, Cambridge.
\newblock


\bibitem{kasparov2010}
{Garry Kasparov}. 2010.
\newblock \showarticletitle{The {Chess} {Master} and the {Computer}}.
\newblock {\em The New York Review of Books\/} {57}, 2 (Feb. 2010), 7.
\newblock


\bibitem{kling1991}
{Rob Kling}. 1991.
\newblock \showarticletitle{Computerization and social transformations}.
\newblock {\em Science, Technology, \& Human Values\/} {16}, 3 (1991),
  342--367.
\newblock


\bibitem{kuijer2018}
{Lenneke Kuijer} {and} {Elisa Giaccardi}. 2018.
\newblock \showarticletitle{Co-performance: {Conceptualizing} the role of
  artificial agency in the design of everyday {Life}}. In {\em Conference on
  {Human} {Factors} in {Computing} {Systems} ({CHI})}. Association for
  Computing Machinery, Montréal, 13.
\newblock


\bibitem{latour1992}
{Bruno Latour}. 1992.
\newblock \showarticletitle{Where are the missing masses, the sociology of
  mundane artefacts}.
\newblock In {\em Shaping {Technology}/{Building} {Society}: {Studies} in
  {Sociotechnical} {Change}}, {Wiebe~E Bijker} {and} {John Law} (Eds.). MIT
  Press, Cambridge, 255--258.
\newblock


\bibitem{ledo2018}
{David Ledo}, {Steven Houben}, {Jo Vermeulen}, {Nicolai Marquadt}, {Lora
  Oehlberg}, {and} {Saul Greenberg}. 2018.
\newblock \showarticletitle{Evaluation strategies for {HCI} toolkit research}.
  In {\em Conference on {Human} {Factors} in {Computing} {Systems} ({CHI})}.
  Association for Computing Machinery, Montréal, 18.
\newblock


\bibitem{leonardi2011}
{Paul~M Leonardi}. 2011.
\newblock \showarticletitle{When flexible routines meet flexible technologies:
  {Affordance}, constraint, and the imbrication of human and material
  agencies}.
\newblock {\em MIS Quarterly\/} {35}, 1 (2011), 147--167.
\newblock


\bibitem{mark2002}
{Gloria Mark}. 2002.
\newblock \showarticletitle{Extreme collaboration}.
\newblock {\it Commun. ACM} {45}, 6 (2002), 89--93.
\newblock


\bibitem{milan2018}
{Matthew Milan}. 2018.
\newblock The {Next} {User} {You} {Design} {For} {Won}’t {Be} {A} {Human}.
\newblock   (Nov. 2018).
\newblock
\showURL{%
\url{https://www.fastcompany.com/90146967/the-next-user-you-design-for-wont-be-a-human}}


\bibitem{morgan1980}
{Gareth Morgan}. 1980.
\newblock \showarticletitle{Paradigms, metaphors, and puzzle solving in
  organization theory}.
\newblock {\em Administrative Science Quarterly\/} {25}, 4 (1980), 605--622.
\newblock


\bibitem{myers2000}
{Brad Myers}, {Scott~E Hudson}, {and} {Randy Pausch}. 2000.
\newblock \showarticletitle{Past, present, and future of user interface
  software tools}.
\newblock {\em ACM Transactions on Computer-Human Interaction\/} {7}, 1 (2000),
  3--28.
\newblock
\showDOI{%
\url{http://dx.doi.org/10.1145/344949.344959}}


\bibitem{nardi1991}
{Bonnie~A Nardi} {and} {James~R Miller}. 1991.
\newblock \showarticletitle{Twinkling lights and nested loops: distributed
  problem solving and spreadsheet development}.
\newblock {\em International Journal of Man-Machine Studies\/} {34}, 2 (1991),
  161--184.
\newblock


\bibitem{olsen2007}
{Dan~R. Olsen}. 2007.
\newblock \showarticletitle{Evaluating {User} {Interface} {Systems}
  {Research}}. In {\em {UIST} {Symposium} on {User} {Interface} {Software} and
  {Technology}}. Association for Computing Machinery, Newport, 251--258.
\newblock
\showDOI{%
\url{http://dx.doi.org/10.1145/1294211.1294256}}


\bibitem{orlikowski2008}
{Wanda~J Orlikowski} {and} {Susan~V Scott}. 2008.
\newblock \showarticletitle{Sociomateriality: challenging the separation of
  technology, work and organization}.
\newblock {\em The Academy of Management Annals\/} {2}, 1 (2008), 433--474.
\newblock


\bibitem{orlikowski2014}
{Wanda~J. Orlikowski} {and} {Susan~V. Scott}. 2014.
\newblock \showarticletitle{What happens when evaluation goes online?
  {Exploring} apparatuses of valuation in the travel sector}.
\newblock {\em Organization Science\/} {25}, 3 (2014), 868--891.
\newblock


\bibitem{parkin2003}
{Kevin~LG Parkin}, {Joel~C Sercel}, {Michael~J Liu}, {and} {Daniel~P
  Thunnissen}. 2003.
\newblock \showarticletitle{Icemaker™: an excel-based environment for
  collaborative design}. In {\em {IEEE} {Aerospace} {Conference}}. IEEE, Big
  Sky, 11.
\newblock


\bibitem{rittel1973}
{Horst~WJ Rittel} {and} {Melvin~M. Webber}. 1973.
\newblock \showarticletitle{Dilemmas in a general theory of planning}.
\newblock {\em Policy Sciences\/} {4}, 2 (1973), 155--169.
\newblock


\bibitem{robertson2009}
{BF Robertson} {and} {DF Radcliffe}. 2009.
\newblock \showarticletitle{Impact of {CAD} tools on creative problem solving
  in engineering design}.
\newblock {\em Computer-Aided Design\/} {41}, 3 (2009), 136--146.
\newblock


\bibitem{schon1983}
{D Schon}. 1983.
\newblock {\em The {Reflective} {Practitioner}: {How} {Professionals} {Think}
  in {Action}}.
\newblock Temple Smith, London.
\newblock


\bibitem{schultze2000}
{Ulrike Schultze}. 2000.
\newblock \showarticletitle{A confessional account of an ethnography about
  knowledge work}.
\newblock {\em MIS Quarterly\/} {24}, 1 (2000), 3--41.
\newblock


\bibitem{shestakofsky2017}
{Benjamin Shestakofsky}. 2017.
\newblock \showarticletitle{Working algorithms: {Software} automation and the
  future of work}.
\newblock {\em Work and Occupations\/} {44}, 4 (2017), 376--423.
\newblock


\bibitem{topoptfig}
{Siemens~PLM Software}. 2017.
\newblock Generative design: {Optimize} shapes to achieve design goals.
\newblock   (2017).
\newblock
\showURL{%
\url{https://solidedge.siemens.com/pt-br/pages/generative-design-solid-edge-ebook/}}


\bibitem{sketchpad}
{Ivan~E. Sutherland}. 1964.
\newblock \showarticletitle{Sketchpad: a man-machine graphical communication
  system}.
\newblock {\em Simulation\/} {2}, 5 (1964), 18.
\newblock


\end{thebibliography}

\end{document}